\documentclass[preprint2]{aastex}

\shorttitle{Multiwavelength Observations of the VHE Blazar  1ES\,2344+514}
\shortauthors{Acciari et al.}

\begin{document}

\title{Multiwavelength Observations of the VHE Blazar 1ES\,2344+514}
%%%%%%
%%%%%%%%%%%%%%%%%%%%%%
%%
\author{
V. A. Acciari\altaffilmark{1},
E. Aliu\altaffilmark{2},
T. Arlen\altaffilmark{3},
T. Aune\altaffilmark{4},
M. Beilicke\altaffilmark{5},
W. Benbow\altaffilmark{1},
D. Boltuch\altaffilmark{2},
V. Bugaev\altaffilmark{5},
A. Cannon\altaffilmark{6},
L. Ciupik\altaffilmark{7},
P. Cogan\altaffilmark{8},
P. Colin\altaffilmark{9},
R. Dickherber\altaffilmark{5},
A. Falcone\altaffilmark{10},
S. J. Fegan\altaffilmark{3},
J. P. Finley\altaffilmark{11}, 
P. Fortin\altaffilmark{12},
L. F. Fortson\altaffilmark{7},
A. Furniss\altaffilmark{4},
D. Gall\altaffilmark{11},
G. H. Gillanders\altaffilmark{13},
J. Grube\altaffilmark{6,7,*}
R. Guenette\altaffilmark{8},
G. Gyuk\altaffilmark{7},
D. Hanna\altaffilmark{8},
J. Holder\altaffilmark{2},
D. Horan\altaffilmark{14},
C. M. Hui\altaffilmark{9},
T. B. Humensky\altaffilmark{15},
A. Imran\altaffilmark{16},
P. Kaaret\altaffilmark{17},
N. Karlsson\altaffilmark{7},
M. Kertzman\altaffilmark{18},
D. Kieda\altaffilmark{9},
J. Kildea\altaffilmark{1},
A. Konopelko\altaffilmark{19},
H. Krawczynski\altaffilmark{5},
F. Krennrich\altaffilmark{16},
M. J. Lang\altaffilmark{13},
S. LeBohec\altaffilmark{9},
G. Maier\altaffilmark{8},
P. Moriarty\altaffilmark{20},
R. Mukherjee\altaffilmark{12},
R. A. Ong\altaffilmark{3},
A. N. Otte\altaffilmark{4},
D. Pandel\altaffilmark{17},
J. S. Perkins\altaffilmark{1},
A. Pichel\altaffilmark{21},
M. Pohl\altaffilmark{16},
J. Quinn\altaffilmark{6},
K. Ragan\altaffilmark{8},
P. T. Reynolds\altaffilmark{22},
H. J. Rose\altaffilmark{23},
M. Schroedter\altaffilmark{16},
G. H. Sembroski\altaffilmark{11},
A. W. Smith\altaffilmark{24},
D. Steele\altaffilmark{7},
S. P. Swordy\altaffilmark{15},
M. Theiling\altaffilmark{1},
J. A. Toner\altaffilmark{13},
A. Varlotta\altaffilmark{11},
V. V. Vassiliev\altaffilmark{3},
S. Vincent\altaffilmark{9},
R. Wagner\altaffilmark{24},
S. P. Wakely\altaffilmark{15},
J. E. Ward\altaffilmark{6},
T. C. Weekes\altaffilmark{1},
A. Weinstein\altaffilmark{3},
T. Weisgarber\altaffilmark{15},
D. A. Williams\altaffilmark{4},
S. Wissel\altaffilmark{15},
M. Wood\altaffilmark{3},
B. Zitzer\altaffilmark{11}
}

\altaffiltext{1}{Fred Lawrence Whipple Observatory, Harvard-Smithsonian Center for Astrophysics, Amado, AZ 85645, USA}
\altaffiltext{2}{Department of Physics and Astronomy and the Bartol Research Institute, University of Delaware, Newark, DE 19716, USA}
\altaffiltext{3}{Department of Physics and Astronomy, University of California, Los Angeles, CA 90095, USA}
\altaffiltext{4}{Santa Cruz Institute for Particle Physics and Department of Physics, University of California, Santa Cruz, CA 95064, USA}
\altaffiltext{5}{Department of Physics, Washington University, St. Louis, MO 63130, USA}
\altaffiltext{6}{School of Physics, University College Dublin, Belfield, Dublin 4, Ireland} 
\altaffiltext{7}{Astronomy Department, Adler Planetarium and Astronomy Museum, Chicago, IL 60605, USA}
\altaffiltext{8}{Physics Department, McGill University, Montreal, QC H3A 2T8, Canada}
\altaffiltext{9}{Physics Department, University of Utah, Salt Lake City, UT 84112, USA}
\altaffiltext{10}{Department of Astronomy and Astrophysics, 525 Davey Lab, Pennsylvania State University, University Park, PA 16802, USA}
\altaffiltext{11}{Department of Physics, Purdue University, West Lafayette, IN 47907, USA }
\altaffiltext{12}{Department of Physics and Astronomy, Barnard College, Columbia University, NY 10027, USA}
\altaffiltext{13}{School of Physics, National University of Ireland, Galway, Ireland}
\altaffiltext{14}{Laboratoire Leprince-Ringuet, Ecole Polytechnique, CNRS/IN2P3, F-91128 Palaiseau, France}
\altaffiltext{15}{Enrico Fermi Institute, University of Chicago, Chicago, IL 60637, USA}
\altaffiltext{16}{Department of Physics and Astronomy, Iowa State University, Ames, IA 50011, USA}
\altaffiltext{17}{Department of Physics and Astronomy, University of Iowa, Van Allen Hall, Iowa City, IA 52242, USA}
\altaffiltext{18}{Department of Physics and Astronomy, DePauw University, Greencastle, IN 46135-0037, USA}
\altaffiltext{19}{Department of Physics, Pittsburg State University, 1701 South Broadway, Pittsburg, KS 66762, USA}
\altaffiltext{20}{Department of Life and Physical Sciences, Galway-Mayo Institute of Technology, Dublin Road, Galway, Ireland}
\altaffiltext{21}{Instituto de Astronomia y Fisica del Espacio, Casilla de Correo 67 - Sucursal 28, (C1428ZAA) Ciudad Aut—noma de Buenos Aires, Argentina}
\altaffiltext{22}{Department of Applied Physics and Instrumentation, Cork Institute of Technology, Bishopstown, Cork, Ireland}
\altaffiltext{23}{School of Physics and Astronomy, University of Leeds, Leeds, LS2 9JT, UK}
\altaffiltext{24}{Argonne National Laboratory, 9700 S. Cass Avenue, Argonne, IL 60439, USA}
\altaffiltext{*}{Corresponding author: jgrube@adlerplanetarium.org}

%%
%%%%%%%%%%%%%%%%%%%%%%%%%%%%%%%%%%%%%%%%%%%%%
%%
\begin{abstract}
Multiwavelength observations of the high-frequency-peaked blazar 1ES\,2344+514 
were performed from 2007 October to 2008 January. The campaign represents the 
first contemporaneous data on the object at very high energy (VHE, E $>$100 GeV) 
$\gamma$-ray, X-ray, and UV energies. Observations with VERITAS in VHE 
$\gamma$-rays yield a strong detection of 20 $\sigma$ with 633 excess events 
in a total exposure of 18.1 hours live-time. A strong VHE $\gamma$-ray flare on 
2007 December 7 is measured at F($>$300 GeV) $= (6.76 \pm 0.62) \times 10^{-11}$ 
ph cm$^{-2}$ s$^{-1}$, corresponding to 48\% of the Crab Nebula flux. Excluding this 
flaring episode, nightly variability at lower fluxes is observed with a 
time-averaged mean of F($>$300 GeV) $= (1.06 \pm 0.09) \times 10^{-11}$ ph 
cm$^{-2}$ s$^{-1}$ (7.6\% of the Crab Nebula flux). The differential photon 
spectrum between 390 GeV and 8.3 TeV for the time-averaged observations excluding 
2007 December 7 is well described by a power law with a photon index of 
$\Gamma = 2.78 \pm 0.09_{\rm{stat}} \pm 0.15_{\rm{syst}}$. On the flaring night of 2007 
December 7 the measured VHE $\gamma$-ray photon index 
was $\Gamma = 2.43 \pm 0.22_{\rm{stat}} \pm 0.15_{\rm{syst}}$. 
Over the full period of 
VERITAS observations contemporaneous X-ray and UV data were taken 
with \emph{Swift} and \emph{RXTE}. The measured 2--10 keV flux ranged by a 
factor of $\sim$7 during the campaign. On 2007 December 8 the highest ever 
observed X-ray flux from 1ES\,2344+514 was measured by \emph{Swift} XRT at a flux 
of F(2--10 keV) $= (6.28 \pm 0.31)\times 10^{-11}$ erg cm$^{-2}$ s$^{-1}$. 
Evidence for a correlation between the 
X-ray flux and VHE $\gamma$-ray flux on nightly time-scales is indicated with 
a Pearson correlation coefficient of $r = 0.60 \pm 0.11$. Contemporaneous 
spectral energy distributions (SEDs) of 1ES\,2344+514 are presented for two distinct 
flux states. A one-zone synchrotron self-Compton (SSC) model describes both SEDs 
using parameters consistent with previous SSC modeling of 1ES\,2344+514 from 
non-contemporaneous observations. 
%%%
\end{abstract}
\section{Introduction}
%%
%%%%%%
The majority of blazars detected at very high energy (VHE, E $>$100 GeV) 
$\gamma$-rays are high-frequency-peaked BL Lac objects (HBLs), with currently 
$\sim$25 HBLs from a total of $\sim$30 VHE blazars.\footnote{The TeVCat catalog of VHE 
$\gamma$-rays sources is available online at: http://tevcat.uchicago.edu} 
The short time-scale variability seen in the broadband spectral energy 
distributions (SEDs) of blazars is explained by highly relativistic plasma 
jets oriented close to the line of sight \citep{Blandford79}. HBLs are 
blazars exhibiting synchrotron radiation peaking typically at UV to 
X-ray energies and a second SED component peaking at GeV to TeV energies. 
Leptonic models \citep{Coppi92,Bottcher02,Krawczynski04} describe 
the high-energy peak as inverse Compton upscattering of low-energy photons 
by electrons, while hadronic models attribute the emission to proton-induced 
cascades, synchrotron radiation by protons, pion production in dense clumps 
of jet plasma, or curvature radiation \citep{Mannheim93,Aharonian00,Mucke01}. 
Detailed multiwavelength studies from optical to $\gamma$-ray energies of the 
temporal and spectral variability of HBLs promise to constrain the physical 
parameters of the underlying particle distributions, particularly the Doppler 
factor and magnetic field strength \citep{Tavecchio98}. 

The HBL 1ES\,2344+514 was first detected in the Einstein Slew Survey 
at X-ray energies (0.2--4 keV) and has a redshift of z $=$ 0.044 
\citep{Elvis92,Perlman96}. VHE $\gamma$-ray emission from 1ES\,2344+514 was 
discovered by the Whipple 10 m telescope in energies $>$350 GeV at a 
6.0 $\sigma$ detection level during a 1 day flare on 1995 December 20 
\citep{Catanese98}. X-ray observations of 1ES\,2344+514 with \emph{BeppoSAX} 
revealed variability in the 2--10 keV flux by a factor of $\sim$2 during a 
7-day period in 1996 December \citep{Giommi00}. 1ES\,2344+514 was later 
observed by \emph{Swift} in 2005 during April, May, and December at a lower 
flux than in 1996 December \citep{Tramacere07}. The X-ray spectrum is reasonably 
well described by an absorbed power law, with photon indices ranging 
from $\Gamma$ $\approx$ 1.8--2.3. Marginal evidence for optical variability 
in 1ES\,2344+514 is seen from the sparse data set of two observations in 1998 
\citep{Nilsson99, Falomo99}, six nights in 2000 \citep{Xie02}, and one 
observation on 2007 January 12 \citep{Gupta08}. In the radio band, VLA  
observations reveal an unusual morphology at arcsec scales for a blazar 
of two compact high brightness emission regions connected via a diffuse halo, 
more reminiscent of a radio galaxy \citep{Rector03,Giroletti04}. High 
resolution VLBA observations show a well-collimated jet extending $\sim$10 pc 
from the core, which then bends 25$^{\circ}$ and broadens into a cone of 
$\sim$35$^{\circ}$ opening angle. In high energy (HE, 0.1$>$E$<$300 GeV) 
$\gamma$-rays, 1ES\,2344+514 is associated with the source 1FGL\,J2347.1+5142 
from the \emph{Fermi} LAT First Source Catalog \citep{Abdo10}, detected at 
a significance of 10.6 $\sigma$ and showing a hard HE $\gamma$-ray spectrum 
of 1.57 $\pm$ 0.17 with no indication of HE $\gamma$-ray flux variability 
during the period of 2008 August to 2009 July. 

Evidence for long-term VHE $\gamma$-ray flux variability in 1ES\,2344+514 is 
seen in observations between 1995 and 2005 by the Whipple 10 m telescope, 
HEGRA, and MAGIC. The Whipple 10 m $\gamma$-ray photon spectrum on the flaring 
night of 1995 December 20 is best-fit by a power law over the energy range 
0.8--9 TeV, with a photon index 
$\Gamma = 2.54 \pm 0.17_{\rm{stat}} \pm 0.07_{\rm{syst}}$ \citep{Schroedter05}. 
The measured flux F($>$1 TeV) $=$ $(3.3 \pm 0.7) \times 10^{-11})$ ph cm$^{-2}$ 
s$^{-1}$ for the flare night is a factor of $1.7 \pm 0.3$ higher than the 
Crab Nebula flux. Excluding 1995 December 20, the time-averaged Whipple 10 m data 
between 1995 and 1997 yielded a marginal 4 $\sigma$ detection at a 
significantly lower flux level, corresponding to $\sim$10\% of the Crab Nebula 
flux \citep{Catanese98}. Further VHE $\gamma$-ray observations of 
1ES\,2344+514 with HEGRA in 1997, 1998, and 2002 resulted in a 4.4 $\sigma$ 
detection in a low flux level state of ($3.3 \pm 1.0$)\% of the Crab Nebula 
flux \citep{Aharonian04}. Observations in 1995 with TACTIC above 1.5 TeV in 
2004 and 2005 yielded a weakly constraining flux upper limit near the Whipple 
10 m detection flux level \citep{Godambe07}. In 2005 August to 2006 January, 
the MAGIC telescope measured a time-averaged photon spectrum between 0.14--5.4 
TeV at a low flux of 5\% of the Crab Nebula flux characterized by a power-law with 
$\Gamma = 2.95 \pm 0.12_{\rm{stat}} \pm 0.20_{\rm{syst}}$ \citep{Albert07}. 
This article presents the first multiwavelength campaign on 
1ES\,2344+514, with contemporaneous UV, X-ray and VHE $\gamma$-ray 
observations over a several month long period. 
%%
%%%%
\section{VERITAS Observations}
%%%%
%%
%%
\begin{figure}[t]
\includegraphics[scale=.38]{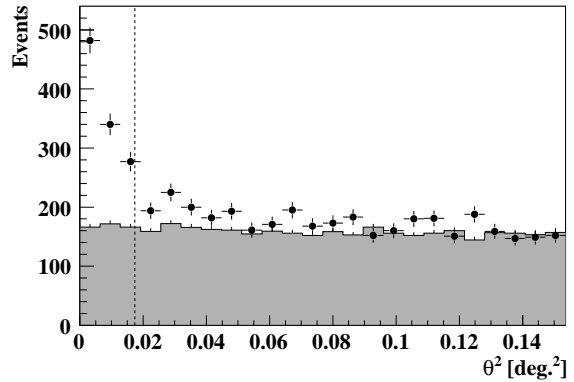}
\caption{The distribution of $\theta^{2}$ for on-source events (points) 
and normalized off-source events (shaded region) from observations of 
1ES\,2344+514. The dashed line represents the applied cut on the 
$\theta^{2}$ parameter.}\label{FigTh2}
\end{figure}
%%%
%%%
VERITAS is an array of four $12\,$m diameter imaging atmospheric Cherenkov 
telescopes located at the Fred Lawrence Whipple Observatory in Southern 
Arizona \citep{Weekes02}. Each VERITAS camera contains 499 pixels 
(0.15$^{\circ}$ diameter), and has a field of view of 3.5$^{\circ}$. VERITAS 
is sensitive over an energy range of $\sim$100 GeV to $\sim$30 TeV with an 
energy resolution of 15--20\%, and an angular resolution (68\% containment) 
of less than 0.14$^\circ$ per event \citep{Acciari08a}. VERITAS can detect 
VHE $\gamma$-ray fluxes of 5\% and 1\% of the Crab Nebula flux 
at the 5 $\sigma$ level in $<$2.5 and $<$50 hours, respectively.

The VERITAS observations of 1ES\,2344+514 presented here were taken on 37 
nights between 2007 October 4 and 2008 January 11. After applying 
quality-selection criteria, the total exposure is 18.1 hours live-time. 
Data-quality selection requires clear atmospheric conditions, based on 
infrared sky temperature measurements, and normal hardware operation. During 
each night, the quality-selected data ranged from 0.3--2.7 hours live-time. 
The zenith angle of observations ranges from 19--48$^{\circ}$, with a mean 
of 28$^{\circ}$. All data were taken during moon-less periods in {\it{wobble}} 
mode with pointings of 0.5$^\circ$ from the blazar alternating from North, 
South, East, and West directions to enable simultaneous background estimation 
\citep{Aharonian01}.

Data reduction followed the methods described in \citet{Acciari08a}. 
Signals in each event are first calibrated \citep{Holder06}, and the images 
are parameterized \citep{Hillas85}. The $\gamma$-ray direction and air shower 
impact parameter on the ground are then reconstructed using stereoscopic 
techniques \citep{Hofmann99,Krawczynski06}. The background of cosmic-rays is 
rejected with a very high efficiency using event-by-event cuts on the arrival 
direction ($\theta^2$), mean scaled width and length, integrated charge 
({\it size}), and location of the image centroids in the camera 
({\it distance}). The cuts applied here are optimized {\it a priori} 
for a source strength of 10\% of the Crab Nebula flux and a similar photon 
index to the Crab Nebula. The energy of each event is reconstructed using 
lookup tables from Monte Carlo simulations of $\gamma$-rays 
\citep{Acciari08a}. Results from two independent VERITAS analysis packages 
\citep{Daniel07} yield results consistent with those presented here.  

A total of 1275 on-source and 4494 off-source events were measured, with an 
on-off normalization of $\alpha$ = 0.143.  This corresponds to a total excess 
of 632 events from the direction of 1ES\,2344+514. The statistical 
significance of this excess is 20.2 standard deviations ($\sigma$). 
Figure \ref{FigTh2} shows the $\theta^{2}$ distribution of the total data set, 
which is the squared angular distance between the reconstructed event direction 
and the nominal source position. The shape of the excess is consistent with a 
simulated point source for VERITAS. 
%%%%%
%%
\section{\emph{RXTE} Observations}
%%%
The X-ray satellite mission \emph{RXTE} \citep{Bradt93} observed 
1ES\,2344+514 between 2007 October 7 and 2008 January 11 (ObsID 93132). 
Data are presented from Proportional Counter Unit 2 of the PCA instrument 
\citep{Jahoda96} since for nearly all observations the other PCUs were not 
in operation. The 52 nightly PCA observations were taken in snapshots ranging 
from 0.11-1.00 hours live-time, and are listed in Table \ref{Tab}. Data reduction 
is performed with the \emph{HEAsoft} 6.5 package. Only the top layer (X1L and X1R) 
signal is used. The data are filtered following the standard criteria advised by 
the NASA Guest Observer Facility. Background data are parameterized with the 
\emph{pcabackest} tool using the most recent model for faint sources. 
The photon spectrum of each observation is extracted using the 
\emph{saextrct} tool. Response matrices are generated using \emph{pcarsp} 
with the latest calibration files. 
%%%
\section{\emph{Swift} Observations}
%%%
\emph{Swift} \citep{Gehrels04} Target of Opportunity (ToO) observations of 
1ES\,2344+514 from 2007 October 27 to 2008 January 1 were triggered by 
the VERITAS detection. The \emph{Swift} data set consists of eight snapshot 
observations of 10--45 minute duration each, as listed in Table \ref{Tab}. 
All \emph{Swift} XRT data \citep{Burrows05} are 
reduced using the \emph{HEAsoft} 6.5 package. Event files are calibrated 
and cleaned following the standard filtering criteria using the 
\emph{xrtpipeline} task and  applying the most recent \emph{Swift} XRT 
calibration files. All data were taken in Photon Counting (PC) mode, with 
grades 0--12 selected over the energy range 0.3--10 keV. Due to photon 
pile-up in the core of the point spread function (PSF) at count rates 
$>$0.5 counts s$^{-1}$ (PC mode), the source events are extracted from an 
annular region with inner radius ranging from 2--6 pixels and an outer 
radius of 30 pixels (47.2 arcsec). Background counts are extracted from a 
40 pixel radius circle in a source-free region. Ancillary response files are 
generated using the \emph{xrtmkarf} task, with corrections applied for the 
PSF losses and CCD defects. The latest response matrix from the XRT calibration 
files is applied. To ensure valid $\chi^{2}$ minimization statistics during 
spectral fitting, the extracted XRT energy spectra are rebinned to contain a 
minimum of 30 counts in each bin.

UVOT observations were taken in the photometric bands of \emph{UVW1}, 
\emph{UVM2}, and \emph{UVW2} \citep{Poole08}. The \emph{uvotsource} tool is 
used to extract counts, correct for coincidence losses, apply background 
subtraction, and calculate the source flux. The standard 5 arcsec radius 
source aperture is used, with a 20 arcsec background region. The source 
fluxes are dereddened using the interstellar extinction curve in 
\citet{Fitzpatrick99}. In the UV filters only a low level of host galaxy 
flux is evident \citep{Tramacere07}, so no corrections are needed. The 
uncertainty in point spread function variations with time is fixed for this 
analysis at 15\%. Observations were taken with just one UV filter during each 
pointing, and only \emph{UVM2} data were taken on multiple nights. Within the 
conservative errors, no variability is seen between the four \emph{UVM2} 
filter observations. 
%%%%
%%%
\section{Spectral Analysis}
%%%
%%
\begin{figure}[t]
\includegraphics[scale=.38]{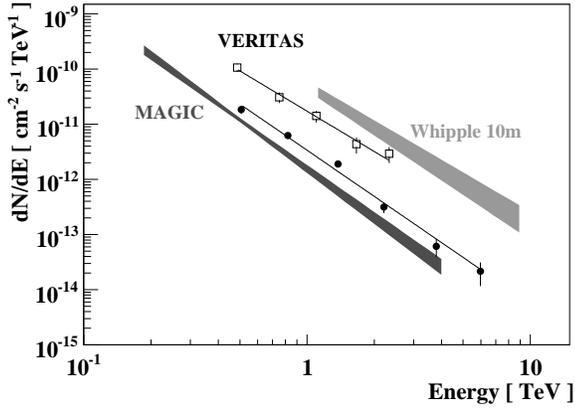}
\caption{Differential photon spectrum of VHE $\gamma$-rays for 1ES\,2344+514. 
The circles represent the time-averaged VERITAS data from 2007 October to 
2008 January, excluding the bright flare on 2007 December 7. The open squares 
represent VERITAS data from the flare night. Shown for comparison are the 68\% 
confidence intervals for the best-fit power law models to 1ES\,2344+514 data 
from MAGIC (dark grey band) in 2005 August to 2006 January \citep{Albert07}, 
and from the Whipple 10 m (light grey band) during a high flux state on 
1995 December 20 \citep{Schroedter05}. 
}\label{FigSpec}
\end{figure}
%%%
%%
\begin{figure}[t]
\includegraphics[scale=.38]{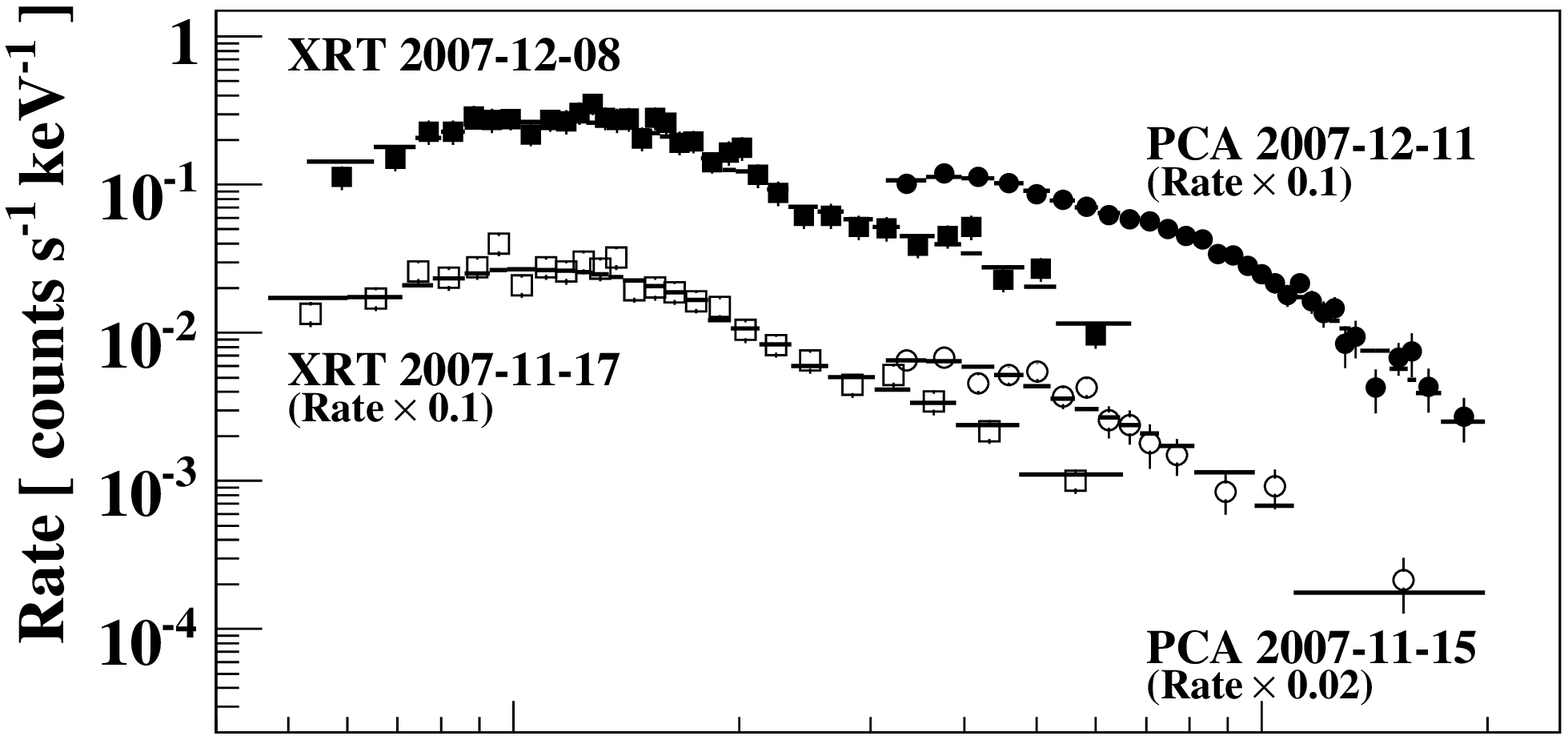}
\includegraphics[scale=.38]{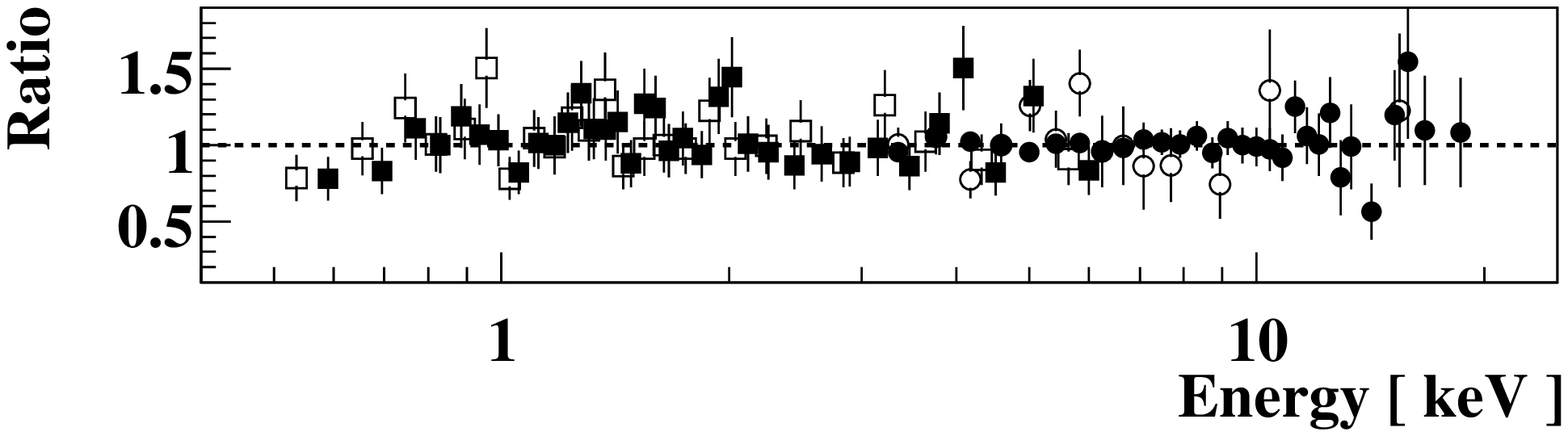}
\caption{
X-ray spectrum of 1ES\,2344+514 from \emph{Swift} XRT and \emph{RXTE} PCA observations. Shown in the top panel are example moderate-flux and high-flux spectra identified in the plot by instrument and observation date. The differential rates are scaled in three cases for viewing purposes. For each energy bin, the statistical significance is $>$3 $\sigma$, and the horizontal line represents the best-fit forward folded absorbed power law model. The bottom panel shows the ratio of the data and model values. 
}\label{FigXspec}
\end{figure}
%%%
%%%%%%%%%%
\begin{table*}
\begin{center} {\footnotesize 
\caption{Best-fit spectral parameters for the \emph{Swift} XRT and \emph{RXTE} PCA data.}\label{Tab}
\begin{tabular}{ccccccc}
\hline
\hline
Date & Start & Exp. & $\Gamma$ & K$_{\rm{1 keV}}$ & $\chi^{2}_{\rm{r}}$/dof & F(2--10 keV) \\
             & (hr:min UT) & (ksec) &  & (10$^{-2}$ ph cm$^{-2}$ s$^{-1}$ keV$^{-1}$) &  & (10$^{-11}$ erg cm$^{-2}$ s$^{-1}$) \\
\hline
\emph{Swift} XRT & &      &                 &                 &         &           \\
2007-10-27 & 10:24 & 2.66 & 2.33 $\pm$ 0.05 & 0.62 $\pm$ 0.03 & 1.11/36 & 0.96 $\pm$ 0.06 \\
2007-11-03 & 03:02 & 1.97 & 2.32 $\pm$ 0.06 & 0.87 $\pm$ 0.04 & 1.02/27 & 1.37 $\pm$ 0.09 \\
2007-11-10 & 05:03 & 0.63 & 2.26 $\pm$ 0.13 & 1.26 $\pm$ 0.10 & 0.69/8  & 2.18 $\pm$ 0.31 \\
2007-11-17 & 02:28 & 1.65 & 2.16 $\pm$ 0.06 & 0.76 $\pm$ 0.04 & 0.71/23 & 1.51 $\pm$ 0.14 \\
2007-11-30 & 03:46 & 1.85 & 2.10 $\pm$ 0.07 & 0.91 $\pm$ 0.05 & 1.51/23 & 1.99 $\pm$ 0.14 \\
2007-12-08 & 03:20 & 2.22 & 1.97 $\pm$ 0.05 & 2.37 $\pm$ 0.10 & 0.77/33 & 6.28 $\pm$ 0.31 \\
2007-12-22 & 04:28 & 2.23 & 2.18 $\pm$ 0.05 & 1.00 $\pm$ 0.04 & 1.12/36 & 1.93 $\pm$ 0.13 \\
2008-01-01 & 00:35 & 1.37 & 2.10 $\pm$ 0.08 & 1.26 $\pm$ 0.07 & 1.07/18 & 2.72 $\pm$ 0.17 \\
\hline
\emph{RXTE} PCA &  &      &           &           &         &   \\ 
2007-10-07 & 06:03 & 2.56 & 2.48 $\pm$ 0.26 & 0.75 $\pm$ 0.35 & 0.52/38 & 0.95 $\pm$ 0.26 \\
2007-10-08 & 06:56 & 2.88 & 2.65 $\pm$ 0.20 & 1.24 $\pm$ 0.38 & 0.58/38 & 1.24 $\pm$ 0.15 \\
2007-10-09 & 06:14 & 2.80 & 2.31 $\pm$ 0.13 & 1.04 $\pm$ 0.23 & 0.54/38 & 1.66 $\pm$ 0.10 \\
2007-10-10 & 05:53 & 2.67 & 2.17 $\pm$ 0.12 & 0.89 $\pm$ 0.19 & 0.47/38 & 1.76 $\pm$ 0.14 \\
2007-10-11 & 03:42 & 2.16 & 2.57 $\pm$ 0.15 & 1.65 $\pm$ 0.40 & 0.42/38 & 1.85 $\pm$ 0.11 \\
2007-10-12 & 03:14 & 0.21 & 2.57 $\pm$ 0.19 & 1.34 $\pm$ 0.41 & 0.66/38 & 1.48 $\pm$ 0.15 \\
2007-10-13 & 04:52 & 2.02 & 2.26 $\pm$ 0.13 & 1.11 $\pm$ 0.24 & 0.63/38 & 1.92 $\pm$ 0.15 \\
2007-10-14 & 02:17 & 1.86 & 2.39 $\pm$ 0.12 & 1.63 $\pm$ 0.33 & 0.62/38 & 2.35 $\pm$ 0.14 \\
2007-10-15 & 03:56 & 1.50 & 2.49 $\pm$ 0.16 & 1.61 $\pm$ 0.41 & 0.53/38 & 2.02 $\pm$ 0.20 \\
2007-10-16 & 04:35 & 1.84 & 2.25 $\pm$ 0.18 & 0.80 $\pm$ 0.27 & 0.53/38 & 1.41 $\pm$ 0.24 \\
2007-10-19 & 07:09 & 3.62 & 2.42 $\pm$ 0.14 & 1.02 $\pm$ 0.23 & 0.51/38 & 1.39 $\pm$ 0.13 \\
2007-10-20 & 08:18 & 3.15 & 2.61 $\pm$ 0.15 & 1.51 $\pm$ 0.35 & 0.59/38 & 1.59 $\pm$ 0.12 \\
2007-10-21 & 07:57 & 3.49 & 2.36 $\pm$ 0.16 & 0.81 $\pm$ 0.23 & 0.56/38 & 1.22 $\pm$ 0.11 \\
2007-11-01 & 03:06 & 2.59 & 2.19 $\pm$ 0.15 & 0.73 $\pm$ 0.21 & 0.79/38 & 1.40 $\pm$ 0.13 \\
2007-11-02 & 05:53 & 1.39 & 2.41 $\pm$ 0.18 & 1.25 $\pm$ 0.37 & 0.62/38 & 1.73 $\pm$ 0.16 \\
2007-11-03 & 07:06 & 2.37 & 2.72 $\pm$ 0.19 & 1.65 $\pm$ 0.48 & 0.53/38 & 1.50 $\pm$ 0.14 \\
2007-11-04 & 04:56 & 2.85 & 2.19 $\pm$ 0.14 & 0.75 $\pm$ 0.20 & 0.50/38 & 1.43 $\pm$ 0.11 \\
2007-11-05 & 06:35 & 0.80 & 2.20 $\pm$ 0.29 & 0.68 $\pm$ 0.40 & 0.64/38 & 1.28 $\pm$ 0.45 \\
2007-11-06 & 04:02 & 2.48 & 2.23 $\pm$ 0.11 & 1.10 $\pm$ 0.21 & 0.82/38 & 2.00 $\pm$ 0.11 \\
2007-11-07 & 06:36 & 0.31 & 2.43 $\pm$ 0.12 & 1.28 $\pm$ 0.25 & 0.49/38 & 1.75 $\pm$ 0.09 \\
2007-11-08 & 07:52 & 2.96 & 2.20 $\pm$ 0.10 & 1.10 $\pm$ 0.18 & 0.57/38 & 2.07 $\pm$ 0.11 \\
2007-11-09 & 07:35 & 2.40 & 2.10 $\pm$ 0.09 & 1.16 $\pm$ 0.17 & 0.59/38 & 2.53 $\pm$ 0.12 \\
2007-11-10 & 03:50 & 2.53 & 2.13 $\pm$ 0.08 & 1.41 $\pm$ 0.18 & 0.33/38 & 2.94 $\pm$ 0.09 \\
2007-11-11 & 01:49 & 2.27 & 2.16 $\pm$ 0.08 & 1.50 $\pm$ 0.20 & 0.63/38 & 3.02 $\pm$ 0.13 \\
2007-11-12 & 04:31 & 2.51 & 2.23 $\pm$ 0.13 & 0.91 $\pm$ 0.21 & 0.54/38 & 1.65 $\pm$ 0.16 \\
2007-11-13 & 02:30 & 2.30 & 2.11 $\pm$ 0.09 & 1.19 $\pm$ 0.18 & 0.71/38 & 2.57 $\pm$ 0.13 \\
2007-11-14 & 03:37 & 2.40 & 2.22 $\pm$ 0.11 & 1.15 $\pm$ 0.21 & 0.55/38 & 2.12 $\pm$ 0.10 \\
2007-11-15 & 03:10 & 2.34 & 2.50 $\pm$ 0.16 & 1.31 $\pm$ 0.33 & 0.60/38 & 1.61 $\pm$ 0.15 \\
2007-11-30 & 02:20 & 0.37 & 2.44 $\pm$ 0.26 & 1.91 $\pm$ 0.81 & 0.74/38 & 2.54 $\pm$ 0.37 \\
2007-12-01 & 01:51 & 0.50 & 2.59 $\pm$ 0.33 & 1.62 $\pm$ 0.84 & 0.38/38 & 1.76 $\pm$ 0.46 \\
2007-12-02 & 05:03 & 0.58 & 2.23 $\pm$ 0.18 & 1.48 $\pm$ 0.43 & 0.72/38 & 2.65 $\pm$ 0.25 \\
2007-12-04 & 02:31 & 0.40 & 2.10 $\pm$ 0.19 & 1.32 $\pm$ 0.42 & 0.61/38 & 2.87 $\pm$ 0.39 \\
2007-12-05 & 02:11 & 1.79 & 2.01 $\pm$ 0.05 & 2.15 $\pm$ 0.19 & 0.81/38 & 5.37 $\pm$ 0.15 \\
2007-12-06 & 03:28 & 1.57 & 2.06 $\pm$ 0.06 & 1.99 $\pm$ 0.22 & 0.69/38 & 4.63 $\pm$ 0.11 \\
2007-12-10 & 02:18 & 2.03 & 1.94 $\pm$ 0.06 & 1.39 $\pm$ 0.15 & 0.90/38 & 3.90 $\pm$ 0.10 \\
2007-12-11 & 02:50 & 1.82 & 1.87 $\pm$ 0.04 & 1.97 $\pm$ 0.15 & 0.62/38 & 6.10 $\pm$ 0.11 \\
2007-12-13 & 03:38 & 3.39 & 1.86 $\pm$ 0.04 & 1.54 $\pm$ 0.10 & 0.67/38 & 4.85 $\pm$ 0.07 \\
2007-12-12 & 05:41 & 1.23 & 1.97 $\pm$ 0.06 & 1.96 $\pm$ 0.21 & 0.46/38 & 5.20 $\pm$ 0.13 \\
2007-12-14 & 04:35 & 3.33 & 2.08 $\pm$ 0.06 & 1.53 $\pm$ 0.15 & 0.65/38 & 3.42 $\pm$ 0.09 \\
2007-12-15 & 04:53 & 0.86 & 2.17 $\pm$ 0.13 & 1.57 $\pm$ 0.33 & 0.91/38 & 3.10 $\pm$ 0.23 \\
2007-12-28 & 03:46 & 1.06 & 2.14 $\pm$ 0.11 & 1.48 $\pm$ 0.28 & 0.72/38 & 3.06 $\pm$ 0.17 \\
2007-12-29 & 04:53 & 1.17 & 1.96 $\pm$ 0.15 & 0.69 $\pm$ 0.22 & 0.47/38 & 1.86 $\pm$ 0.23 \\
2007-12-30 & 04:25 & 0.88 & 1.91 $\pm$ 0.09 & 1.40 $\pm$ 0.22 & 0.56/38 & 4.04 $\pm$ 0.22 \\
2008-01-01 & 01:52 & 0.86 & 2.06 $\pm$ 0.11 & 1.43 $\pm$ 0.27 & 0.85/38 & 3.30 $\pm$ 0.15 \\
2008-01-02 & 02:26 & 2.66 & 2.12 $\pm$ 0.07 & 1.40 $\pm$ 0.17 & 0.53/38 & 2.96 $\pm$ 0.11 \\
2008-01-03 & 03:34 & 2.91 & 1.97 $\pm$ 0.06 & 1.23 $\pm$ 0.13 & 0.67/38 & 3.26 $\pm$ 0.10 \\
2008-01-04 & 01:53 & 1.71 & 2.06 $\pm$ 0.13 & 0.87 $\pm$ 0.20 & 0.65/38 & 2.03 $\pm$ 0.12 \\
2008-01-05 & 02:35 & 2.83 & 1.97 $\pm$ 0.07 & 1.02 $\pm$ 0.13 & 0.51/38 & 2.71 $\pm$ 0.11 \\
2008-01-06 & 02:08 & 2.75 & 2.31 $\pm$ 0.12 & 1.08 $\pm$ 0.22 & 0.41/38 & 1.76 $\pm$ 0.11 \\
2008-01-07 & 03:48 & 1.98 & 2.33 $\pm$ 0.14 & 1.16 $\pm$ 0.27 & 0.68/38 & 1.83 $\pm$ 0.12 \\
2008-01-10 & 02:23 & 1.50 & 2.25 $\pm$ 0.10 & 1.78 $\pm$ 0.29 & 0.52/38 & 3.12 $\pm$ 0.14 \\
2008-01-11 & 03:24 & 2.77 & 2.12 $\pm$ 0.07 & 1.53 $\pm$ 0.17 & 0.74/38 & 3.25 $\pm$ 0.10 \\
\hline
\end{tabular} }
\end{center}
\end{table*}
%%%%%%%
%%%%%%
%% \begin{deluxetable}{ccccccc}
%% %%\tabletypesize{\scriptsize}
%% %%\rotate
%% %%\tablewidth{200pt}
%% %%\tablenum{}
%% %%\tablecolumns{}
%% \tablehead{\colhead{Date} & \colhead{Start} & \colhead{Exp.} & \colhead{$\Gamma$} & \colhead{K$_{1 keV}$} & \colhead{$\chi^{2}_{\rm{r}}$/dof} & \colhead{F(2--10 keV)} &
%% \nodata & \colhead{(hr:min)} & \colhead{(ksec)} & \nodata & \colhead{(10$^{-3}$ cm$^{-2}$ s$^{-1}$ keV$^{-1}$)} & \nodata & \colhead{(10$^{-11}$ erg cm$^{-2}$ s$^{-1}$)} 
%%  }
%%%%
The VERITAS differential photon spectrum over the energy range of 390 GeV 
to 8.3 TeV for the time-averaged data excluding a strong flaring night on 
2007 December 7 is shown in Figure \ref{FigSpec}. A power law 
model fit to the data is best described by the form 
$d\rm{N}/d\rm{E} = \rm{F}_{\circ} \cdot (\rm{E}/1 TeV)^{-\Gamma}$ yielding 
a $\chi^{2}$/dof of 8.39/4 for a flux normalization constant of 
$\rm{F}_{\circ} = (3.38 \pm 0.23_{\rm{stat}} \pm 0.70_{\rm{syst}}) 
\times 10^{-12}$ ph cm$^{-2}$ s$^{-1}$ TeV$^{-1}$ and a photon index of 
$\Gamma = 2.78 \pm 0.09_{\rm{stat}} \pm 0.15_{\rm{syst}}$. A fit 
to a power law with exponential cutoff model, which contains one additional 
degree of freedom, does not yield a significantly better fit. On 2007 
December 7 the weather conditions were affected by moving cloud coverage, so 
the observations do not pass the standard data-quality criteria. However, 
online analysis showed that the source was in the most active state yet 
observed. We therefore include these data and increase the systematic 
uncertainties in flux and spectral shape. Conservative systematic 
uncertainties in the flux of 30\% and photon index of 0.2 are determined, 
based on the results of observations of the Crab Nebula taken in similarly 
variable conditions. The VHE $\gamma$-ray spectrum from the high flux night 
is well described by a power law ($\chi^{2}$/dof of 1.27/3) with 
a slightly harder photon index $\Gamma = 2.43 \pm 0.22_{\rm{stat}} 
\pm 0.2_{\rm{syst}}$, and a higher flux normalization of 
$\rm{F}_{\circ} = (17.3 \pm 1.92_{\rm{stat}} \pm 5.19_{\rm{syst}}) 
\times 10^{-12}$ ph cm$^{-2}$ s$^{-1}$ TeV$^{-1}$. 

The best-fit results presented here are tested with a dedicated Monte-Carlo 
study, which explores the uncertainty ranges for the two measured spectra. In 
the Monte-Carlo study photons are selected from a power-law distribution matching 
the measured distribution convolved with the effective area 
as a function of energy.  The selected photon energies are then smeared with 
a Gaussian corresponding to the energy resolution of $\sim$15\% at these energies. 
The number of photons selected is matched to the observed number of photons with a 
Poisson scatter. Five million synthetic photon spectra are created using similar 
binning to the original spectra, and the best-fit power law index and normalization 
parameters are extracted for each synthetic spectrum providing dense coverage of 
the parameter space. The Monte-Carlo mean value and one dimensional standard 
deviation of the power law parameters are consistent with the best-fit 
parameters of the measured spectra. For the time-averaged spectrum 
excluding the flaring night of 2007 December 7, 
the Monte-Carlo study yields a mean value and 1-$\sigma$ error for the 
flux normalization of $\rm{F}_{\circ,\rm{MC}} = (3.35 \pm 0.30) \times 10^{-12}$ 
ph cm$^{-2}$ s$^{-1}$ TeV$^{-1}$ and for the the photon index of 
$\Gamma_{\rm{MC}} = 2.76 \pm 0.15$. For the high flux 2007 December 7 spectrum, the 
Monte-Carlo study yields $\rm{F}_{\circ,MC} = (17.3 \pm 2.67) \times 10^{-12}$ 
ph cm$^{-2}$ s$^{-1}$ TeV$^{-1}$ and $\Gamma_{\rm{MC}} = 2.40 \pm 0.28$.

Previous VHE $\gamma$-ray observations of 1ES\,2344+514 
with the MAGIC telescope between 2005 August and 2006 January measured a 
constant flux at a slightly lower level than the time-averaged (flare excluded) 
flux measured here, and a consistent photon index of 
$\Gamma = 2.95 \pm 0.12_{\rm{stat.}} \pm 0.2_{\rm{syst.}}$ \citep{Albert07}. 
The high flux VHE $\gamma$-ray photon spectrum measured by the Whipple 10 m 
on 1995 December 20 with a photon index of 
$\Gamma = 2.54 \pm 0.17_{\rm{stat}} \pm 0.07_{\rm{syst}}$ \citep{Schroedter05} 
is marginally harder than the low flux spectrum and agrees with the high 
flux spectrum measured here. The VHE observed $\gamma$-ray spectral shape and 
intensity are modified by absorption from pair production interactions with 
the infrared component of the extragalactic background light (EBL) \citep{Gould67}. 
Using the redshift z $=$ 0.044 for 1ES\,2344+514, we calculate a de-absorbed 
spectrum based on the EBL model of \citet{Franceschini08}, yielding intrinsic 
photon indices for the VERITAS low and high flux states of 
$\Gamma_{\rm{int,low}} \approx 2.5$ and $\Gamma_{\rm{int,high}} \approx 2.1$, 
respectively. 
 
X-ray spectral analysis of the \emph{Swift} XRT and \emph{RXTE} PCA data 
is performed with \emph{XSPEC} 12.4. An absorbed power law model, including 
the \emph{phabs} model for the photoelectric absorption, is fit to 
each spectrum. First, a joint fit of the eight \emph{Swift} XRT over the 
energy range 0.4--10 keV using a tied column density N$_{\rm{H}}$ and each 
spectrum having a varying photon index and normalization is compared to a 
joint fit with a fixed Galactic column density 
N$_{\rm{H,Gal}}$ of $1.50 \cdot 10^{21}$ cm$^{-2}$ \citep{Kalberla05}. The joint 
fit with tied N$_{\rm{H}}$ yields a best-fit 
N$_{\rm{H}}$ of $(2.06 \pm 0.13) \cdot 10^{21}$ cm$^{-2}$, with a 
reduced $\chi^{2}$ of 1.03 for 203 degrees of freedom (dof) 
and chance probability for larger chi-squared statistics (null hypothesis) of 37\%. 
The alternative joint fit with the fixed Galactic column density N$_{\rm{H,Gal}}$ 
yields a higher reduced $\chi^{2}$ of 1.13 for 204 dof and null hypothesis of 9.2\%. 
The best-fit N$_{\rm{H}}$ is then used in the absorbed power law model for the 
\emph{RXTE} PCA spectra, which are mostly unaffected by absorption due to the 
3--20 keV energy range, and for all \emph{Swift} XRT results presented here. 
Figure \ref{FigXspec} shows \emph{Swift} XRT (0.4--10 keV) and 
\emph{RXTE} PCA (3--20 keV) spectra from example moderate-flux and 
high-flux observations. The ratio in each energy bin of the data to the 
absorbed power law model is shown for the example spectra, with the full list 
of reduced $\chi^{2}$ values per dof for each observation listed in Table 
\ref{Tab}. The model describes well the spectra, yielding an average chance 
probability for larger chi-squared statistics of  37\% and 51\% for the 
\emph{Swift} XRT and \emph{RXTE} PCA data, respectively. 
The measured \emph{Swift} XRT 0.4--10 keV photon indices range from 
$\Gamma = 2.33 \pm 0.05$ to $\Gamma = 1.97 \pm 0.05$ for observations differing 
in 2--10 keV integrated flux by a factor of 6.54 $\pm$ 0.52. These results 
are consistent with the \emph{BeppoSAX} 0.5--10 keV photon indices of 
$\Gamma$ $\approx$ 1.8--2.3 at similar flux levels for an absorbed power law 
model with fixed Galactic column density \citep{Giommi00}. Detailed results 
from the nightly \emph{RXTE} PCA and \emph{Swift} XRT spectra focusing on the 
connection between the significant X-ray spectral and flux variability are 
presented in the following section. 
%%%
\section{Light Curves}
%%%
\begin{figure*}
\includegraphics[scale=.82]{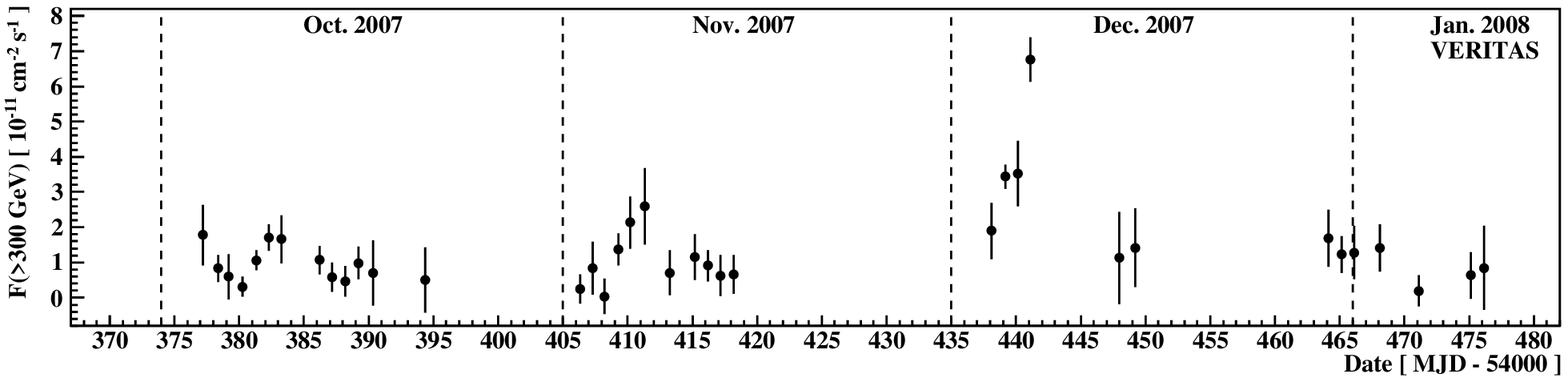}
\includegraphics[scale=.82]{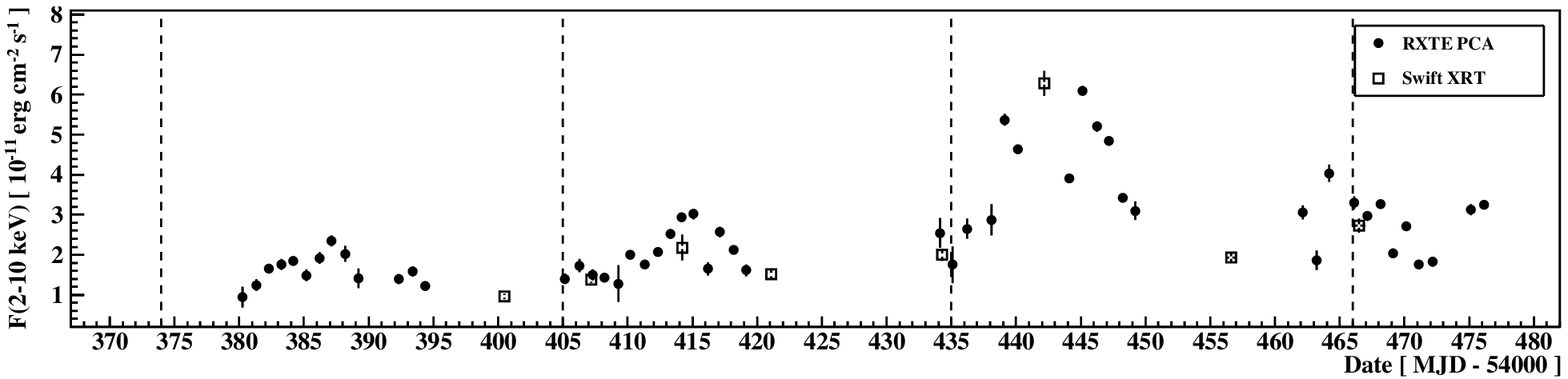}
\includegraphics[scale=.82]{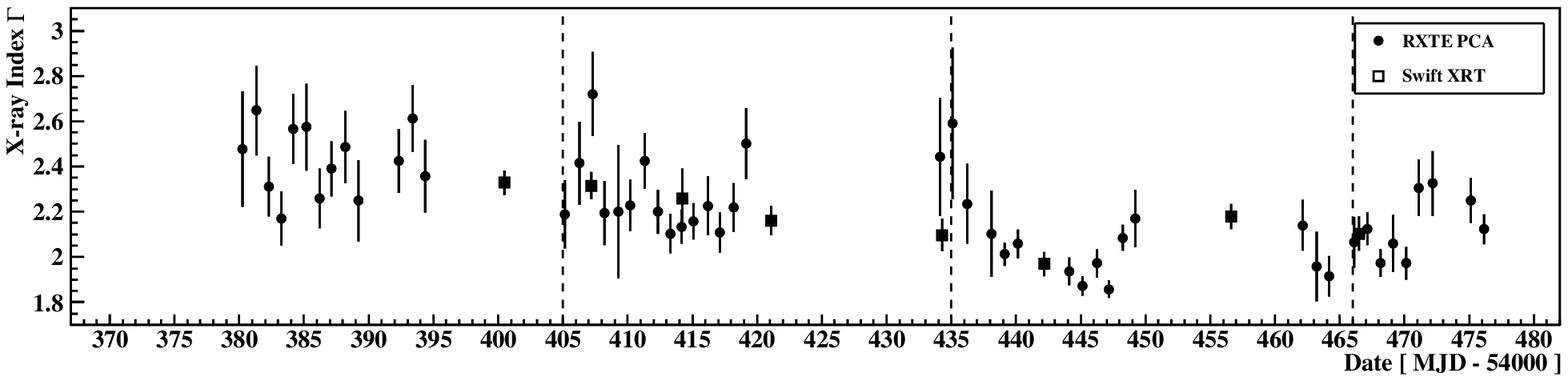}
\caption{VHE $\gamma$-ray and X-ray light curve of 1ES\,2344+514. Shown 
in the top panel are VERITAS F($>$300 GeV) nightly fluxes. The middle 
panel shows the 2--10 keV fluxes from observations with \emph{RXTE} PCA 
(circles) and \emph{Swift} XRT (open squares). The bottom panel shows X-ray 
photon indices from the best-fit absorbed power law model.
}\label{FigLC}
\end{figure*}
%%%
%%%%
\begin{figure}
\includegraphics[scale=.38]{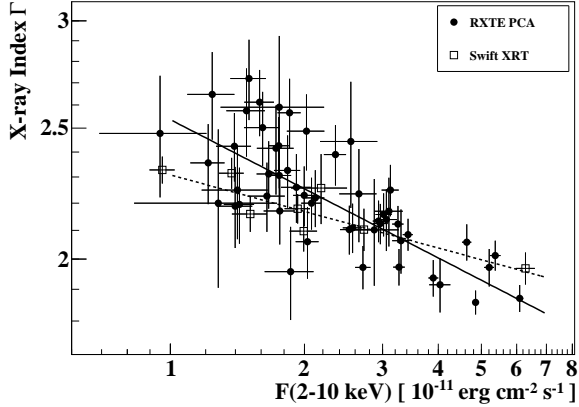}
\caption{X-ray photon index versus 2-10 keV flux from \emph{RXTE} PCA 
(circles) and \emph{Swift} XRT (open squares) data on a log-log scale. The  
best-fit power law for the \emph{RXTE} PCA (3--20 keV) data is represented 
by a solid line, and for \emph{Swift} XRT (0.4--10 keV) 
with a dashed line.}\label{FigXcorr}
\end{figure}
%%%
%%%%
\begin{figure}
\includegraphics[scale=.38]{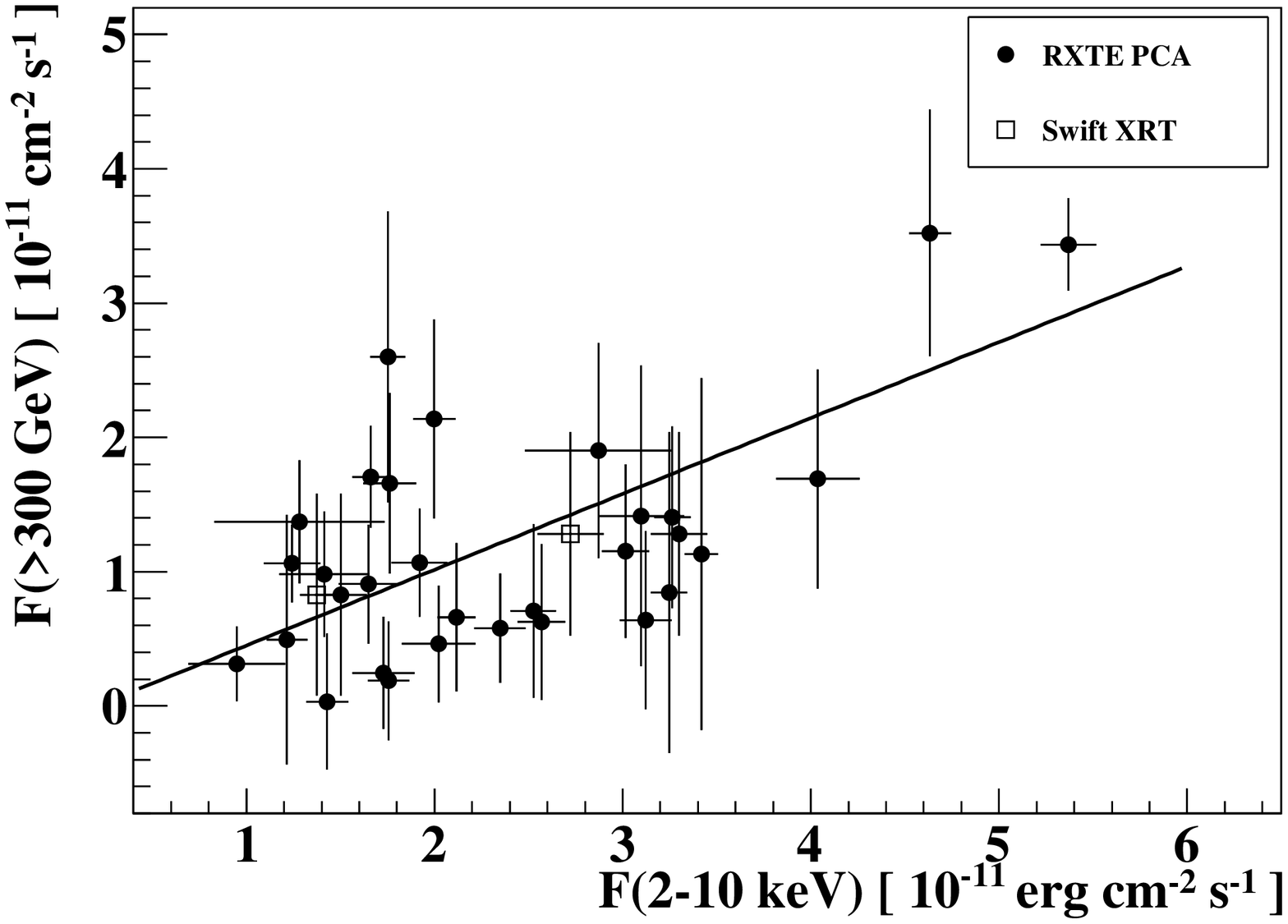}
\caption{VERITAS $\gamma$-ray flux F($>$300 GeV) versus X-ray 2--10 keV flux 
from nights with \emph{RXTE} PCA (circles) and \emph{Swift} XRT (open squares) 
data. A linear fit to the data is represented by 
the solid line.}\label{FigFcorr}
\end{figure}
%%%
%%%
The nightly VHE $\gamma$-ray and X-ray light curve of 1ES\,2344+514 from 
VERITAS, \emph{RXTE} PCA, and \emph{Swift} XRT is shown in Figure \ref{FigLC}. 
A strong VHE $\gamma$-ray flare is seen on 2007 December 7 (54441.12 MJD) 
at an integrated flux F($>$300 GeV) $= (6.8 \pm 0.6) \times 10^{-11}$ 
cm$^{-2}$ s$^{-1}$, corresponding to 48\% of the Crab Nebula flux. The 
measured increase in flux of a factor of 1.9 $\pm$ 0.5 between 
the previous night and the flare night shows the first clear evidence of 
$\sim$day time-scale VHE $\gamma$-ray variability from 1ES\,2344+514 since the 
Whipple 10 m discovery of VHE $\gamma$-ray emission in 1995 
\citep{Catanese98}. Excluding the 2007 December 7 flaring event, the average 
F(E$>$300 GeV) is $(1.1 \pm 0.1) \times 10^{-11}$ ph cm$^{-2}$ s$^{-1}$, 
corresponding to 7.6\% of the Crab Nebula flux. A fit to a constant flux 
excluding the flaring night is rejected with a chance probability of 
$7.2 \times 10^{-8}$, indicating significant low-level variability. A 
measure of the integrated level of flux variability is the fractional 
root-mean-square (rms) variability amplitude F$_{\rm{var}}$ 
\citep{Vaughan03}. For the full VERITAS data set of nightly exposures a high 
level of variability F$_{\rm{var}} = (75 \pm 10$)\% is implied. Excluding 
the flare night, a still significant F$_{\rm{var}} = (34 \pm 16$)\% is 
determined. Searches for short-term flux variations within each of the nightly 
observations showed no significant variability. 

Figure \ref{FigLC} (lower panels) shows the 2--10 keV flux and photon index 
$\Gamma$ measured over 3--20 keV from \emph{RXTE} PCA and 0.4--10 keV from 
\emph{Swift} XRT data. The X-ray flux is shown to be highly variable 
throughout the campaign, with F$_{\rm{var}} = (51 \pm 1$)\%. In 2007 December, 
large amplitude flaring is evident with flux doubling time-scales of 
$\sim$1 day. A 2--10 keV flux of $(6.3 \pm 0.3) \times 10^{-11}$ erg 
cm$^{-2}$ s$^{-1}$ is seen from the Swift XRT data on 2007 December 8, 
representing the highest X-ray flux ever measured for 1ES\,2344+514. 
Analysis of all subsequent X-ray data of 1ES\,2344+514, which currently 
consists solely of \emph{Swift} XRT observations, show a 2--10 keV flux 
consistent with the lowest flux measurements presented here. Detailed results 
from the more recent \emph{Swift} XRT data set are beyond the scope of this 
publication, but automated count rate light curves are publicly 
available.\footnote{The \emph{Swift} XRT Monitoring of \emph{Fermi} LAT Sources 
of Interest is available online at: http://www.swift.psu.edu/monitoring/}

X-ray spectral variability between flaring nights 
is investigated by plotting the X-ray photon index $\Gamma$ versus the 
2--10 keV flux, shown in Figure \ref{FigXcorr}. A logarithmic correlation of 
decreasing photon index with increasing flux is suggested from a power law 
fit to the \emph{RXTE} PCA and \emph{Swift} XRT data with a $\chi^{2}$/dof 
probability of 0.14 and 0.70 compared to 7.9 $\times 10^{-20}$ and 3.3 
$\times 10^{-4}$ for a constant photon index, respectively. Due to the lack 
of significant curvature in the X-ray spectrum, the peak energy E$_{\rm{p}}$ 
for a $\nu$F($\nu$) SED representation is largely unconstrained. A comparison 
of the photon indices from the \emph{Swift} XRT and \emph{RXTE} PCA spectra 
restricted to the overlapping energy range of 3--10 keV is limited by the 
low number of energy bins above 3 keV, so a study of the systematically higher 
\emph{RXTE} PCA (3--20 keV) photon indices compared with the 
\emph{Swift} XRT (0.4--10 keV) photon indices at similar flux levels is inconclusive.
Furthermore, the lack of purely simultaneous \emph{RXTE} PCA and \emph{Swift} XRT 
data does not allow for a detailed study of joint fits from 0.4--20 keV. 
Even during the bright flaring events in 2007 December, the \emph{RXTE} PCA spectra 
from 3--20 keV show no sign of curvature, with measured photon indices of $\sim$1.9. 
For these high-flux states the implied peak energy E$_{\rm{p}} \leq 10$ keV agrees 
with the estimated peak energies derived from \emph{BeppoSAX} observations in 
1998 \citep{Giommi00}. 

Figure \ref{FigFcorr} shows the VERITAS flux F($>$300 GeV) versus 
\emph{RXTE} PCA and \emph{Swift} XRT 2--10 keV fluxes for nights with 
observations in both energy bands. A Pearson coefficient of $r = 0.60 \pm 0.11$ 
is calculated for the VHE $\gamma$-ray to X-ray flux points, suggestive of 
correlated variability. The best-fit linear model yields a slope of 
$0.56 \pm 0.08$ with a fit probability of 0.14. For the highest measured X-ray 
fluxes in 2007 December the weather conditions at the VERITAS site were poor, 
which excludes these nightly fluxes from figure \ref{FigFcorr}. Due to the 
sparse data sampling, a search for lags between the VHE $\gamma$-ray and 
X-ray emission is not investigated here. 
%%%
%%%
\section{Discussion}
%%%%
\begin{figure*}
\includegraphics[scale=.8]{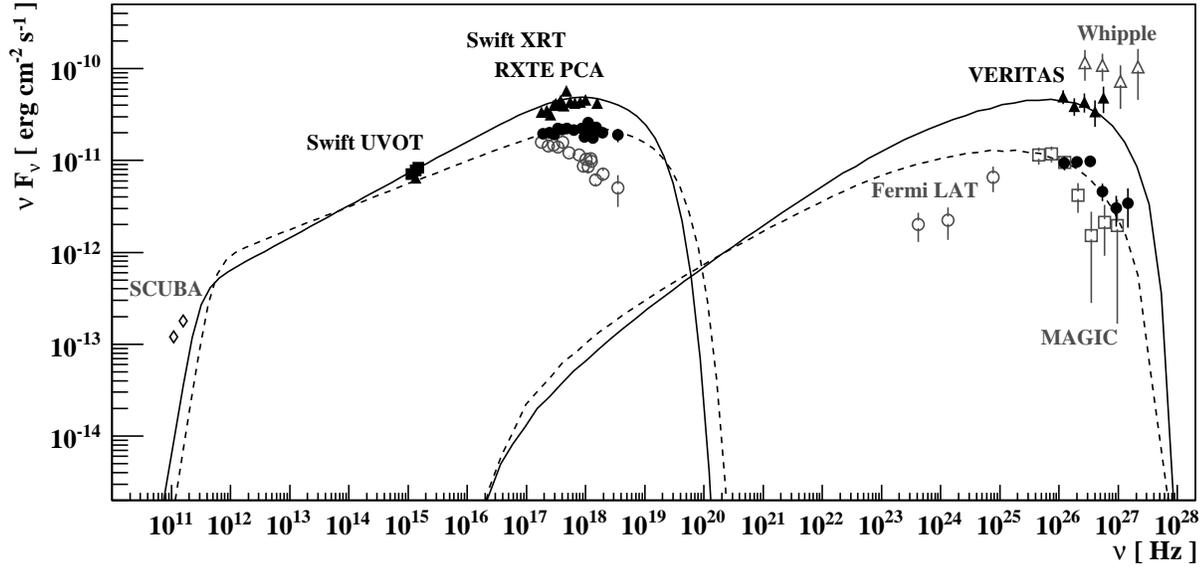}
\caption{Spectral energy distribution of 1ES\,2344+514. The high flux spectrum 
(triangles) are from VERITAS data on 2007 December 7 and \emph{Swift} UVOT and 
XRT data on 2007 December 8. The time-averaged VERITAS spectrum and moderate 
flux level X-ray spectrum from \emph{Swift} XRT and \emph{RXTE} PCA 
data on 2008 January 1 are represented by circles. An example low flux X-ray 
spectrum (open circles) is from 2007 November 3. Non-contemporaneous VHE 
$\gamma$-ray spectra are shown from MAGIC data in 2005 August--December 
(open squares) and from Whipple 10 m data on the flare night of 1995 December 20 
(open triangles). All VHE $\gamma$-ray spectra are corrected for absorption 
by the extragalactic background light \citep{Franceschini08}. Archival millimeter 
fluxes (open diamonds) are from SCUBA data \citep{Stevens99}. The 
non-contemporaneous \emph{Fermi} LAT spectrum from the 1FGL catalog is represented 
by open circles \citep{Abdo10}. The broadband curves are from synchrotron 
self-Compton (SSC) modeling to the contemporaneous data 
\citep{Krawczynski04}. }\label{FigSED}
\end{figure*}
%%%
%%%%
The broadband spectral energy distribution (SED) of high-frequency-peaked 
BL Lac objects (HBLs) is often modeled within a synchrotron self-Compton 
(SSC) framework. Simple one-zone SSC models typically predict correlated 
X-ray and VHE $\gamma$-ray variability from the underlying electron 
distribution. Figure \ref{FigSED} shows a one-zone SSC model calculation 
\citep{Krawczynski04} to the SED of 1ES\,2344+514 for illustration purposes 
in two flux states. A high flux SED is constructed from the VERITAS photon 
spectrum in 2007 December 7 and \emph{Swift} UVOT and XRT data on 2007 December 
8. The time-averaged (flare excluded) VERITAS data is combined with 
\emph{Swift} and \emph{RXTE} PCA spectra on 2008 January 1 when the object 
was at a moderate X-ray flux state with respect to the full campaign. 
Shown also in Figure \ref{FigSED} are the non-contemporaneous MAGIC data 
from 2005 in a low flux state and the flaring Whipple 10 m data from 1995 
December 20. Also shown for reference is the \emph{Fermi} LAT spectrum from 
2008 August to 2009 July in a non-variable low flux state. Each model curve 
for the broadband data contemporaneous with the VERITAS data represents the 
intrinsic source radiation, and all VHE $\gamma$-ray spectra are 
corrected for absorption by the extragalactic background light \citep{Franceschini08}. 
The SSC model input parameters for the low (high, when different) states are: a Doppler 
factor $\delta =$ 13 (20), magnetic field $=$ 0.09 G (0.03 G), emission radius $=$ 
10$^{16}$ cm, electron density $=$ 0.05 ergs cm$^{-3}$, and a broken power law 
electron spectrum with low-energies between 10$^{8}$ eV and 10$^{11.3}$ eV 
(10$^{11.4}$ eV) with spectral index n$_{1} =$ 2.5 (2.3), and above the cutoff 
energy with index n$_{1} =$ 3.2 and highest energy of 10$^{12}$ eV.

The light crossing time $\tau = \rm{R}/(c \times \delta) \approx 5$ hours for the high 
flux state with a Doppler factor $\delta = 20$ is consistent with the $\sim$hour time-scale 
X-ray variability seen from 1ES\,2344+514 with \emph{BeppoSAX} on 1996 December 7 
\citep{Giommi00}. Further hourly X-ray variability in 1ES\,2344+514 could not be confirmed 
in this work due to the short ($<$hour) exposures with \emph{Swift} and \emph{RXTE} and 
moderate event statistics. In addition to the low and high flux state modeling from data in 
2007 October to 2008 January, the same SSC model was applied to non-contemporaneous low and 
high flux states derived from \emph{BeppoSAX}, MAGIC, and Whipple 10 m spectra 
\citep{Albert07}. A reasonable agreement between model parameters, such as Doppler factors 
$\sim$10--20 and a magnetic field $\sim$0.03--0.3 G, is evident for both the 
low and high flux states with the archival 1ES\,2344+514 data, and compared with one-zone 
SSC modeling of similar moderate-flux level flares in the HBLs 1ES\,1959+650 
\citep{Tagliaferri08}, 1ES\,0806+524 \citep{Acciari09a}, and 1ES\,1101-213 
\citep{Aharonian07}. A detailed study of the SSC model parameter space (e.g. 
\citep{Tavecchio98}) is not pursued here due to sparseness in the SED coverage and 
degeneracy between model parameters. The simple one-zone SSC modeling presented here 
adequately describes the data, however further optical, X-ray, and VHE $\gamma$-ray 
observations of 1ES\,2344+514 simultaneous with \emph{Fermi} LAT promise to offer stronger 
constraints to emission models.
%%%%
%%%
\section{Conclusion}
%%%%
In this paper, a $\sim$4 month long VHE $\gamma$-ray and X-ray observing campaign 
of 1ES\,2344+514 revealed significant flux variability on $\sim$1 day time-scales. 
In particular, flux doubling between nights is evident in both the X-ray and VHE 
$\gamma$-ray bands in 2007 December, with the highest ever X-ray flux measured from 
1ES\,2344+514 on 2007 December 8. Evidence of correlated nightly X-ray flux 
and VHE $\gamma$-ray flux variability is shown, as generally found in the 
well-studied HBLs Mrk 501 \citep{Krawczynski00}, Mrk 421 \citep{Fossati08}, 
and 1ES\,1959+650 \citep{Krawczynski04}, however so-called orphan flares of increased 
flux in one energy band and not the other have also been observed (e.g. 
\citep{Blazejowski05,Krawczynski04}). For these campaigns sparse temporal 
sampling and insufficient VHE sensitivity to short time-scale variability limited 
any statements on a linear or quadratic relationship between VHE and X-ray 
flux expected under various SSC scenarios during rising or decaying flares 
\citep{Fossati08}. The 1ES\,2344+514 data presented here represents the first contemporaneous
multiwavelength campaign of the object from UV to VHE $\gamma$-rays, but is similarly 
restricted by sparse temporal sampling and broadband coverage in the spectral energy 
distributions. Applying a simple one-zone SSC model reasonably describes the data, and is 
consistent with past modeling of the object and other HBLs. Further optical, X-ray, 
and VHE $\gamma$-ray observations of 1ES\,2344+514 simultaneous with \emph{Fermi} LAT 
in the HE $\gamma$-ray band promise to offer stronger constraints to emission models and 
high sensitivity broadband measurements of high-flux state flaring events. 
%%%%
%%%
\acknowledgments
%%%
This research is supported by grants from the U.S. Department of Energy, the U.S. 
National Science Foundation, and the Smithsonian Institution, by NSERC in Canada, 
by STFC in the UK and by Science Foundation Ireland. Special acknowledgment to 
the Swift and RXTE teams for the support of these observations.
%%%%
%%%%%%

\end{document}